\documentclass{article}
\usepackage{graphicx,amssymb,amsmath}

\begin{document}

\title{Study of a Narrow $\pi^+ \pi^-$ Peak at about 755 MeV/$c^2$ in 
$\bar{p} n \rightarrow 2 \pi^+ 3 \pi^-$ Annihilation at Rest}


\author{
  Mario Gaspero \\  \\
  \em {Dipartimento di Fisica, Sapienza Universit\`a di Roma, and }\\
 \em {Istituto Nazionale di Fisica Nucleare, Sezione di Roma 1 }\\
 \em {Piazzale Aldo Moro 2, I-00185, Rome, Italy } \\
 \texttt{mario.gaspero@roma1.infn.it}
}

\date{}

\maketitle


\mbox{~~~~~~~~~~~~~~~~~~~Scalar} mesons

\begin{abstract}
A narrow peak in the $\pi^+ \pi^-$ mass distribution was seen by the 
Rome-Syracuse Collaboration in $\bar{p} n \rightarrow 2 \pi^+ 3 \pi^-$
annihilation at rest in 1970\@. It was ignored for 40 years. 
The reanalysis of this peak finds that it has the mass 
$757.4 \pm 2.8_{\rm stat} \pm 1.2_{\rm sys}$
MeV/$c^2$ and a width consistent with the experimental resolution.
The evidence of the peak is 5.2 standard deviations.
The peak is generated in $( 1.03 \pm 0.21_{\rm stat} \pm 0.21_{\rm sys})\%$
of the $\bar{p} n$ annihilations at rest.
No spin analysis is possible with the statistics of the experiment but there
are arguments suggesting that it has $J^P = 0^+$\@.
\end{abstract}

\section{Introduction}

\mbox{~~~~In 1970,} the Rome-Syracuse Collaboration (RSC) studied the 
branching ratio of the decay $\omega \rightarrow \pi^+ \pi^-$ \cite{RmSy70}
using the data of the $\bar{p} n$ annihilations at rest collected in the 30'' 
BNL bubble chamber.
The analysis measured the upper limit 4.3\% at 95\% confidence level and found
an unexpected result: the $\pi^+ \pi^-$ mass distribution of 1496 
annihilations at rest
\begin{equation}
\bar{p} n \rightarrow 2 \pi^+ 3 \pi^- 
\label{pn5pi}
\end{equation}
had a narrow peak at about 755 MeV/$c^2$\@.
This distribution is shown in Fig.~1a.

A $\chi^2$ fit of this distribution found that the peak had a significance of 
about 4.5 standard deviations (SD) and a width lower than the experimental
resolution.
No relation was found between the $\pi^+ \pi^-$ and other angular and mass 
distributions.
These facts suggested to the RSC that the peak was generated by a 
fluctuation.

\begin{figure}
\begin{center}
\includegraphics[height=0.25\textheight]{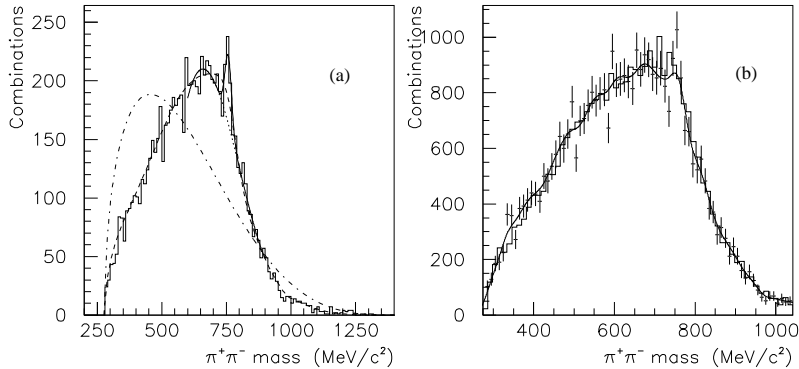}
\end{center}
\caption{
(a) The $\pi^+ \pi^-$ mass distibution of the 
$\bar{p} n \rightarrow 2 \pi^+ 3 \pi^-$ annihilations at rest measured by the
Rome-Syracuse Collaboration.
The dash-dotted line is the phase space prediction; 
the dashed line is the prediction of the analysis of Ref.\ \cite{Gaspero93}; 
the solid line is the prediction of the fit F reported in Table~I;
the dotted line is the polinomial background in the same fit.
(b) The $\pi^+ \pi^-$ mass distibution of the 
$\bar{n} p \rightarrow 3 \pi^+ 2 \pi^-$ annihilations at low momenta measured 
by the OBELIX Collaboration;
the crosses are the Rome-Syracuse data renormalized to the OBELIX combinations;
the solid line is the convolution of the Rome-Syracuse data with a Gaussian 
having the resolution $\mbox{FWHM} = 37.0$ MeV/$c^2$.
}
\end{figure}

The properties of the annihilations \eqref{pn5pi} were undestood at the 
beginning of 1990s \cite{Gaspero92,Gaspero93}.
A reanalysis of the RSC data proved that this reaction is dominated by the 
channel $\bar{p} n \rightarrow f_0(1370) \pi^-$ followed by the $f_0(1370)$ 
decay into $\rho(770)^0 \rho(770)^0$ and $S_w S_w$, where $S_w$ indicates the 
$\pi^+ \pi^-$ $I = 0$ S-wave interaction.
The prediction of Ref.\ \cite{Gaspero93} are shown by
the dashed curves in Fig.~1a.
It fits very well the experimental data, with the
exception of the interval 700-800 MeV/$c^2$\@.
The parametrization of $S_w$ used for obtaining this curve is shown
in Fig.~2.

\begin{figure}
\begin{center}
\includegraphics[height=0.25\textheight]{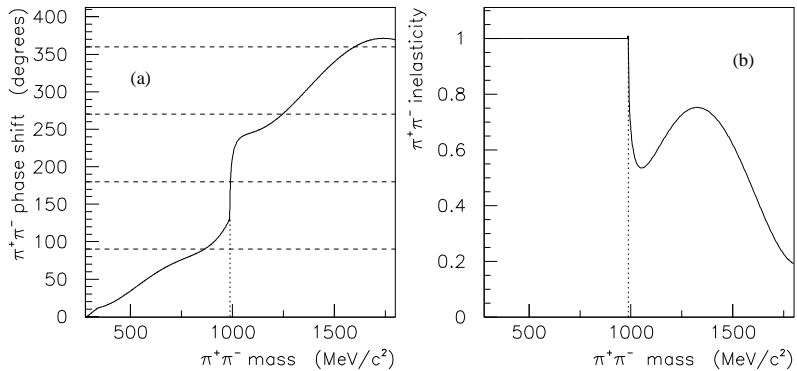}
\end{center}
\caption{
The parametrization RPOH4/5 of the $\pi^+ \pi^-$ $I = 0$ S-wave interaction 
used in Ref.\ \cite{Gaspero93}\@.
The dotted line shows the $K^+ K^-$ threshold.
(a) Phase-shifts.
(b) Inelasticities.
}
\end{figure}

A year after the communication of the preliminary results of this analysis
\cite{Gaspero92}, the OBELIX Collaboration presented the preliminary results 
of the analysis of the charge conjugate annihilation  at low momenta 
\cite{OBELIX93}
\begin{equation}
\bar{n} p \rightarrow 3 \pi^+ 2 \pi^- .
\label{np5pi}
\end{equation}

The $\pi^+ \pi^-$ mass distribution of OBELIX is 
shown in Fig.~1b, togheter with the RSC distribution normalized to the
OBELIX combinations (crosses).
This figure shows that the $\pi^+ \pi^-$ distributions of both experiments 
are in agreement with the only exception of the peak at 
755 MeV/$c^2$ that is not shown by the OBELIX data.
At that time, the absence of the peak in the OBELIX data confirmed the opinion
that the RSC peak was a fluctuation.

Recently, Troyan et al.\ claimed to have observed several narrow
$\pi^+ \pi^-$ peaks in the reaction $n p \rightarrow n p \pi^+ \pi^-$ with 
neutrons of 5.2 GeV/$c$ 
\cite{Troyan}\@.
One of these peak has a mass close to 755 MeV/$c^2$\@.
The coincidence of the mass of the RSC peak with that of Troyan 
et al.\ suggested to reanalyse the RSC data.

\section{Reanalysis of the Rome-Syracuse distribution}
\mbox{~~~~The} $\pi^+ \pi^-$ mas distribution of the RSC has been fitted in 
the mass interval 600 - 900 MeV/$c^2$ with the parametrization
\begin{equation}
D(m) = P_3(m) + N S(m) ,
\label{Dm}
\end{equation}
where $P_3(m)$ is a third degree polinomial, $S(m)$ is the signal
parametrization normalized to one, and $N$ is the number of combinations in
the peak.

The fits have been carried out either with the $\chi^2$ or with the unbinned 
maximum likelihood\footnote{
The maximum likelihood fits were carried out by using the known values of the
$\pi^+ \pi^-$ masses, which were maintained under the support 
of punched cards and afterwards copied manually.}
 methods.
Three signal parametrizations have been used:
\begin{itemize} 
\item[(a)] Gaussian with the resolution $\sigma$ free.
\item[(b)] Gaussian with the resolution fixed at $\sigma = 12.74$ MeV/$c^2$.
           (It correspond to the $\mbox{FWHM} = 30$ MeV/$c^2$ 
           estimated by the RSC in Ref.\ \cite{RmSy70}.)
\item[(c)] The double Gaussian
           \begin{equation}
           S(m) = \frac{f_1}{\sqrt{2 \pi} \, \sigma_1} \,
           e^{-\frac{\displaystyle (M-m)^2}{\displaystyle 2 \sigma_1^2}} +
           \frac{f_2}{\sqrt{2 \pi} \, \sigma_2} \,
           e^{-\frac{\displaystyle (M-m)^2}{\displaystyle 2 \sigma_2^2}} ,
           \label{Gauss2}
           \end{equation}
           with $f_1 = 0.7483$, $f_2 = 0.2517$, $\sigma_1 = 13.24$ MeV/$c^2$,
           and $\sigma_2 = 32.45$ MeV/$c^2$.
           These number have been obtained by fitting the error distribution of
           the Rome subsample.
           (665 events.)
           This function has ${\rm FWHM} = 33.6$ MeV/$c^2$\@.
\end{itemize}

\begin{table}
\caption{
Results of the $\chi^2$ fits of the distribution shown in Fig.~1a
in the interval $600 \leq m \leq 900$ MeV/$c^2$\@.
Fit~A is the fit with $\sigma$ free.
Fit~B is the fit with $\sigma$ fixed at 12.74 MeV/$c^2$.
Fit~C is the fit with the resolution function (\ref{Gauss2}).
$\chi^2_0$ is the $\chi^2$ of the fit with a third degree polynomial without
the signal.
The last row reports the evaluation of the peak significance.
}
\begin{center}
\begin{tabular}{lccc}
\hline
Fit                    &       A         &      B          &           C   \\
\hline
$N$ (combinations)   & $152 \pm 33$    & $188 \pm 37 $   & $220 \pm 45$  \\
$M$ (MeV/$c^2$)    & $755.4 \pm 1.9$ & $756.3 \pm 2.5$ & $756.3 \pm 2.7$ \\
$\sigma$ (MeV/$c^2$) & $7.7 \pm 1.7$   &    12.74        &               \\
$\chi^2$               &     14.7        &     19.6        &    20.5       \\
d.o.f\                 &     24          &      25         &    25         \\
$\chi^2$/d.o.f.\       &     0.61        &     0.78        &    0.82       \\
$\chi^2_0$             &     44.2        &     44.2        &   44.2        \\
$N / \Delta N$     &      4.6        &      5.1        &    4.9        \\
\hline
\end{tabular}
\end{center}
\label{table1}
\end{table}

The results of the fits with the $\chi^2$ and with the maximum likelihood
methods are reported respectively in Table~I and 
in Table~II\@.
These fits prove that the significance of the peak is about 5 SD
when the resolution is fixed using the signals (b) and (c)\@.
In particular, the fits with the maximum likelihood and the parametrizations 
(b) and (c) have both the significance of 5.2 SD\@.
They prove also that the evidence of the peak was not understood in 1970s 
because nobody of the RSC realised that the use of the known experimental 
resolution would have improved the significance above five SD!

\begin{table}
\caption{Results of the unbinned maximum likelihood fits in the interval 
$600 \leq m \leq 900$ MeV/$c^2$\@.
Fit~D is the fit with $\sigma$ free.
Fit~E is the fit with $\sigma = 12.74$ MeV/$c^2$.
Fit~F is the fit with the resolution function (\ref{Gauss2}).
The last  rows report the evaluation of the peak significance.
}
\begin{center}
\begin{tabular}{lccc}
\hline
Fit                    &     D            &      E          &   F \\
\hline
$N$ (combinations)   & $147 \pm 35 $    & $186 \pm 36$  &  $224 \pm 43$  \\
$M$ (MeV/$c^2$)    & $756.7 \pm 1.9$  & $757.4 \pm 2.6$ & $757.4 \pm 2.8$\\
$\sigma$ (MeV/$c^2$) & $ 8.1 \pm 2.1$   &      12.74      & \\
$N / \Delta N$     &      4.2         &     5.2         &    5.2 \\
\hline
\end{tabular}
\end{center}
\label{table2}
\end{table}

The best fit is F, because it takes into account the tails of the error 
distribution and is made with the maximum likelihood method.
Figure~1a shows the prediction of this fit (solid lines) and of the polynomial
background of the same fit (dotted line)\@.

\section{Discussion of the difference between the Rome-Syracuse 
         and the OBELIX data}

\mbox{~~~~The absence} of the peak in the OBELIX $\pi^+ \pi^-$ mass 
distributions can have two explanations:
(i)  the RSC peak was a fluctuation; 
(ii) the OBELIX Collaboration did not see the peak because the
     detector had a resolution poorer than our bubble chamber experiment.

The explanation (i) seems improbable because the unbinned maximum likelihood 
fits with the signals parametrizations (b) and (c) have the significance 5.2 
SD\@.
The explanation (ii) is possible because the OBELIX detector measured
the charged pion tracks after they had traveled through the liquid hydrogen of
the target and the material of the cryogenic vessel.

An indication that the OBELIX Collaboration could not have seen the peak 
because a poorer resolution is given by smoothing the RSC distribution.
The solid line in Fig.~1b shows the convolution obtained by substituting a 
Gaussian having $\mbox{FWHM} = 37.0$ MeV/$c^2$ to all the $\pi^+ \pi^-$ masses 
of the RSC sample.
It estimates the $\pi^+ \pi^-$ mass distribution if the RSC data had had the
resolution $\mbox{FWHM} = 50$ MeV/$c^2$ instead of 33.6 MeV/$c^2$\@.
(In fact,  $50^2 - 33.6^2 = 37.0^2$.)
This curve does not show any peak and is perfectly compatible with the OBELIX 
distribution.

There is a more convincing argument that supports the explanation (ii)\@.
The $\pi^+ \pi^-$ mass distribution is given by
\[ D(m) = \int |A(m;\{x\})|^2 \, d \Phi(\{x\}) =  \, R(m) \Phi(m) \, \]
where $A(m;\{x\})$ is the amplitude of reaction \eqref{pn5pi} that depends on
the $\pi^+ \pi^-$ mass $m$ and on a set $\{x\}$ of other kinematical 
variables, 
$d \Phi(\{x\})$ is the element of the phase space volume,
$\Phi(m)$ is the phase space distribution,
and $R(m)$ is the value of $|A(m;\{x\})|^2$ mediated on the variables 
$\{x\}$\@.
$R(m)$ is the ratio between the experimental distribution and the phase space
(REDPS)
\[ R(m) = \frac{\int |A(m;\{x\})|^2 \, d \Phi(\{x\})}{\int d \Phi(\{x\})}
  = \frac{D(m)}{\Phi(m)} . \]

Figure~3a shows this ratio for the RSC data.
It has two maxima: one at about 755 MeV/$c^2$, and the other at about 1250
 MeV/$c^2$\@.
The first maximum is generated by the peak under study, the second is probably
due to the production of the $f_2(1270)$ meson.
This ratio shows also an ankle at the $K \bar{K}$ threshold and does not show
any evidence for the debated $\sigma(600)$ meson.
Fig~3b shows the REDPS of the OBELIX data.
It has the same behaviour shown by  the REDPS of the RSC, and confirm the 
peaking at about 755 MeV/$c^2$. 
But its maximum is wider than that of the RSC and is not reproduced by the 
prediction of fit F.

\begin{figure}
\begin{center}
\includegraphics[height=0.25\textheight]{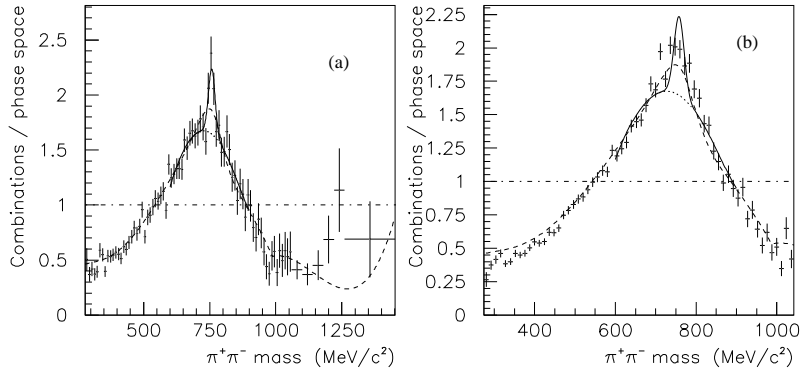}
\end{center}
\caption{
The ratios between the $\pi^+ \pi^-$ mass distibution and the phase space 
predictions.
(a) Ratio of the Rome-Syracuse data.
(b) Ratio of the OBELIX data. 
In both diagrams the dashed line is the prediction of the fit made in Ref.\
\cite{Gaspero93}, 
the solid line is the prediction of the unbinned maximum likelihood fit of the
RSC data with the formulae \eqref{Dm} and \eqref{Gauss2},
and the dotted line is the polinomial background estimated by the same fit.
}
\end{figure}

\section{Quantum numbers of the peak}

\mbox{~~~~The} REDPS distributions prove that there is an amplitude  
peaking at about 755 MeV/$c^2$\@.
This peaking cannot be generated by a reflection because no significative 
correlations were found between the $\pi^+ \pi^-$ peak and other mass and 
angular distributions.
The low mass suggests that it cannot have $J > 1$\@.
Furthermore, the REDPS peak is too narrow for being produced by the 
$\rho(770)^0$ resonance and its mass is 25 MeV/$c^2$ far from that of the 
$\omega$ meson.
Lastly, the peak cannot be due to the $\rho(770)^0 - \omega$ interference.
In fact, the annihilations \eqref{pn5pi} 
are dominated by the S-wave initial states \cite{Gaspero93,Rm74}\@.
Because the G-parity conservation, the channel 
$\bar{p} n \rightarrow \rho(770)^0 \pi^+ 2 \pi^-$ is generated by the 
$^1S_0$ state, while the channel $\bar{p} n \rightarrow \omega \pi^+ 2 \pi^-$
is generated by the $^3S_1$ state.
These states cannot interfere because they are not coherent.

Therefore, the best hypothesis is that the peak is generated by the 
$J^P = 0^+$ interaction.
It could be generated by an unknown scalar meson or could be a property of the
$S_w$ interaction that was not reproduced in the previous analysis 
\cite{Gaspero92,Gaspero93}\@.
In fact, the phase-shits parametrization shown in Fig.~2a is below $90^\circ$ 
till up 862 MeV/$c^2$\@.
This hypothesis sugests that the peak could be reproduced by forcing the phase
shift parametrization to pass at $90^\circ$ at 757.4 MeV/$^2$.
Unfortunately, the events of the RSC are too low for allowing a convincing 
study of the effects of the variations of the $I = 0$ S-wave amplitude.

\section{Conclusions}

\mbox{~~~~The} fit F measures the following parameters for the narrow 
$\pi^+ \pi^-$ peak shown in Fig.~1a
\begin{eqnarray}
M & = & 757.4 \pm 2.8_{\rm stat} \pm 1.2_{\rm sys} \mbox{ MeV/$c^2$}, 
\nonumber \\
\Gamma & < & 30 \mbox{ MeV/$c^2$} , \nonumber \\
N & = & 224 \pm 43_{\rm stat} \pm 45_{\rm sys} \mbox{ combinations}, \label{Np}
\end{eqnarray}
where the systematic error has been evaluated by using the mean of the 
difference between the parameters found in fit F and those measured in the 
other five fits.
The mass $M$ is close to the values measured by Troyan et al.\ \cite{Troyan}.

There is no evidence for double peak production in the same event.
Therefore, one evaluates from \eqref{Np} that the fraction of the 
channel $P(757) \pi^+ 2 \pi^-$ in the $\bar{p} n \rightarrow 2 \pi^* 3 \pi^-$
is
\[ \frac{N}{1496} = ( 15.0 \pm 2.9_{\rm stat} \pm 3.0_{\rm sys})\% . \]
Since the frequency of the annihilation \eqref{pn5pi} at rest is 
$F(\bar{p} n \rightarrow 2 \pi^+ 3 \pi^-) = (6.9 \pm 0.5)\%$ \cite{Gaspero93}, the frequency of the peak is 
\[ F[P(757) \pi^+ 2 \pi^-] = \frac{N}{1496} \, 
F(\bar{p} n \rightarrow 2 \pi^+ 3 \pi^-) = 
      ( 1.03 \pm 0.21_{\rm stat} \pm 0.21_{\rm sys})\% . \]

\section*{Acknowledgments}
\mbox{~~~~I would} like to thank A. Batosi and F. Batosi who did the boring 
work of the manual copying of the $2 \pi$ masses.
I would also like to thank Dr.\ P. Palazzi who has renewed my interest on
this signal.
Lastly, I thank my colleague Dr.\ Fabio Ferrarotto who has helped me to 
revise this paper.

\end{document}